%
%

\input harvmac.tex


\def\np#1#2#3{Nucl. Phys. {\bf B#1} (#2) #3}
\def\pl#1#2#3{Phys. Lett. {\bf #1B} (#2) #3}

\def\prep#1#2#3{Phys. Rep. {\bf #1} (#2) #3}

\def\ap#1#2#3{Ann. Phys. {\bf #1} (#2) #3}
\def\mpl#1#2#3{Mod. Phys. Lett. {\bf A#1} (#2) #3}

\nref\towns{P. K. Townsend, ``The eleven dimensional supermembrane
revisited'', \pl{350}{1995}{184}, hep-th/9501068}%
\nref\witten{E. Witten, ``String theory dynamics in various dimensions'',
\np{443}{1995}{85}, hep-th/9503124}%
\nref\schwarz{J. H. Schwarz, ``An $SL(2,\bZ)$ multiplet of type II
superstrings'', \pl{360}{1995}{13}, hep-th/9508143; 
``Superstring dualities'', hep-th/9509148;
``The power of M theory'', \pl{367}{1996}{97}, hep-th/9510086; 
``M theory extensions of T duality'', hep-th/9601077}%
\nref\hw{P. Ho\v rava and E. Witten, ``Heterotic and type I string
dynamics from eleven dimensions'', \np{460}{1996}{506}, hep-th/9510209}%
\nref\memred{M. J. Duff, P. S. Howe, T. Inami and K. S. Stelle,
``Superstrings in D=10 from supermembranes in D=11'',
\pl{191}{1987}{70}}%
\nref\dfromm{P. K. Townsend, ``D-branes from M-branes'', hep-th/9512062}%
\nref\schmid{C. Schmidhuber, ``D-brane actions'', hep-th/9601003}%
\nref\us{O. Aharony, J. Sonnenschein and S. Yankielowicz,
``Interactions of strings and D-branes from M theory'', hep-th/9603009}%
\nref\aspinwall{P. S. Aspinwall, ``Some relationships between dualities in
string theory'', hep-th/9508154}%
\nref\memmem{P. K. Townsend, ``String-membrane duality in seven
dimensions'', \pl{354}{1995}{247}, hep-th/9504095}%
\nref\hulltowns{C. M. Hull and P. K. Townsend, ``Unity of superstring
dualities'', \np{438}{1995}{109}, hep-th/9410167}%
\nref\km{D. Kutasov and E. Martinec, ``New principles for
string/membrane unification'', hep-th/9602049; D. Kutasov, E. Martinec
and M. O'Loughlin, ``Vacua of M theory and $N=2$ strings'',
hep-th/9603116}%
\nref\dvv{R. Dijkgraaf, E. Verlinde and H. Verlinde, ``BPS spectrum of
the fivebrane and black hole entropy'', hep-th/9603126; ``BPS
quantization of the fivebrane'', hep-th/9604055}%
\nref\newhw{P. Ho\v rava and E. Witten, ``Eleven dimensional supergravity
on a manifold with a boundary'', hep-th/9603142}%
\nref\tdual{A. Giveon, M. Porrati and E. Rabinovici, ``Target space
duality in string theory'', \prep{244}{1994}{77}, hep-th/9401139}%
\nref\bst{E. Bergshoeff, E. Sezgin and P. K. Townsend,
``Supermembranes and eleven dimensional supergravity'',
\pl{189}{1987}{75}; ``Properties of the eleven dimensional
supermembrane theory'', \ap{185}{1988}{330}}%
\nref\tdualII{M. Dine, P. Huet and N. Seiberg, ``Large and small
radius in string theory'', \np{322}{1989}{301}; J. Dai, R. G. Leigh
and J. Polchinski, ``New connections between string theories'',
\mpl{4}{1989}{2073}}%
\nref\sugra{E. Bergshoeff, C. Hull and T. Ortin, ``Duality in the type
II superstring effective action'', \np{451}{1995}{547},
hep-th/9504081}%
\nref\tsen{A. Sen, ``T duality of p-branes'', hep-th/9512203}%
\nref\dyonic{J. M. Izquierdo, N. D. Lambert, G. Papadopoulos and
P. K. Townsend, ``Dyonic membranes'', \np{460}{1996}{560},
hep-th/9508177}%
\nref\witorbs{E. Witten, ``Fivebranes and M theory on an orbifold'',
hep-th/9512219}%
\nref\shenker{S. H. Shenker, ``Another length scale in string
theory?'', hep-th/9509132}%
\nref\newsen{A. Sen, ``Duality and orbifolds'', hep-th/9604070}%
\nref\hethet{M. J. Duff, R. Minasian and E. Witten, ``Evidence for
heterotic/heterotic duality'', hep-th/9601036}%
\nref\ftheory{C. Vafa, ``Evidence for F-theory'', hep-th/9602022; D. R.
Morrison and C. Vafa, ``Compactifications of F-theory on Calabi-Yau
threefolds - I,II'', hep-th/9602114, hep-th/9603161; E. Witten,
``Phase transitions in M theory and F theory'', hep-th/9603150}%
\nref\dmorbs{K. Dasgupta and S. Mukhi, ``Orbifolds of M theory'',
hep-th/9512196}%

\def\bZ{{\bf Z}}
\def\bS{{\bf S}}
\def\bT{{\bf T}}

\def\dg{{\det(G_{ij})}}


\Title{hep-th/9604103, TAUP-2332-96}
{\vbox{\centerline{String theory dualities from M theory}}}

\bigskip

\centerline{\bf Ofer Aharony\foot{Work supported in part by the 
US-Israel Binational Science Foundation, by GIF -- the German-Israeli
Foundation for Scientific Research, and by the Israel Academy of 
Science.}$^,$\foot{Work supported in part by the Clore Scholars Programme.
Address after Sept. 1, 1996 : Department of Physics and Astronomy, Rutgers
University, Piscataway, NJ 08855-0849. 
E-mail address : oferah@post.tau.ac.il.}}
\vglue .5cm
\centerline{School of Physics and Astronomy}
\centerline{Beverly and Raymond Sackler Faculty of Exact Sciences}
\centerline{Tel--Aviv University}
\centerline{Ramat--Aviv, Tel--Aviv 69978, Israel}

\bigskip\bigskip
We analyze how string theory dualities may be described in M
theory. T dualities arise from scalar-vector dualities in the
worldvolume of the membrane of M theory. ``Electric-magnetic''
dualities arise from a duality transformation in M theory
compactified on a 3-torus, which takes the
membrane into a fivebrane wrapped around the 3-torus.

\noindent

\Date{4/96}

\newsec{Introduction}

The conjecture of the existence of an 11 dimensional M theory has led
to a better understanding of many non-perturbative effects in string
theory \refs{\towns,\witten,\schwarz,\hw}. The fundamental
formulation of this theory is not yet known, but many of its
properties may be
derived just from the fact that its low-energy limit is
11 dimensional supergravity. The type IIA string theory and the
heterotic $E_8 \times E_8$ string theory are described
by compactifying M theory on $\bS^1$ \refs{\towns,\witten} and 
$\bS^1 / \bZ_2$ \hw,
respectively. The other consistent 
string theories, the type IIB and the $SO(32)$
theories, can be reached from these by T duality transformations, so
they can only be straightforwardly described by M theory when they are
compactified on a circle \schwarz. All $p$-brane states (for $p \leq 6$) of
these string theories may be identified in M theory by starting with
a membrane and a fivebrane in 11 dimensions
\refs{\towns,\witten,\schwarz}, and their actions (or at
least the field content of their worldvolume theories) may also be
derived from the action (field content) of the membrane and the
fivebrane \refs{\memred,\dfromm,\schmid}. 
Classically, the interactions of these $p$-branes which can
be seen in weakly coupled string theory may also be derived from
simple interactions of the membranes and fivebranes in 11 dimensions \us.

Since M theory is supposed to unify all string theories, it should be
possible to understand the origin of all string theory dualities within M
theory. The simplest string theory dualities, 
the T dualities on a circle, are in some sense
trivially incorporated into M theory since they serve as the M theory
``definitions'' of the type IIB and $SO(32)$ string theories.
We will discuss in section 2 exactly
how the type IIB string action arises
from M theory by an appropriate worldvolume duality transformation of
the membrane.
Other dualities, such as the $SL(2,\bZ)$ duality of the type IIB
string theory, have a simple geometrical origin in M theory
\refs{\schwarz,\aspinwall}. Most of
this paper is devoted to a discussion of
the third type of dualities, which appear to
be related to electric-magnetic duality in M theory. These include,
for instance, the various string-string dualities in 6 dimensions. 
So far the formulation of this electric-magnetic
duality in M theory has not been clear, since it does
not seem to exist in the low-energy 11 dimensional
supergravity theory, due to the existence
of the $C \wedge G \wedge G$ term in the action (where $C$ is the
3-form field
of 11 dimensional supergravity and $G=dC$ is its field strength). 

We would like to propose that the proper setting for electric-magnetic
duality in M theory is 8 dimensions, where one can naturally define a
membrane-membrane duality as originally proposed by Townsend \memmem.
One argument supporting this is that all known string theory dualities
above
8 dimensions \foot{Except for the $SL(2,\bZ)$ duality of the type IIB
string which has a geometrical origin in M theory as mentioned above.}
transform one string theory into another, and, therefore, they
do not necessarily correspond to any symmetry of M theory. Only in 8
dimensions do we have symmetries (in particular, T dualities) which
transform a string theory into itself, and these must indeed be
symmetries also of M theory. Another justification for 
this point of view comes from looking at the U duality groups of
the supergravity theories we get by toroidally
reducing 11 dimensional supergravity \hulltowns. Above 8 dimensions, 
all these
groups have a natural interpretation in M theory, either as
``complex structure'' deformations (for the $SL(2,\bZ)$ group in 9
dimensions) or as a parity transformation (which is the same parity
transformation used by Ho\v rava and Witten in M theory \hw). 
In 8 dimensions, the U duality group is expected to be 
$SL(3,\bZ) \times SL(2,\bZ)$ \hulltowns. The first factor has
an obvious geometrical interpretation in M theory, but the second
does not. As we will show, the $\tau \to -1/\tau$ transformation in
this $SL(2,\bZ)$ group takes the membrane of M theory into a
fivebrane wrapped around $\bT^3$. 
Upon further compactification and orbifolding, this transformation
gives rise to all of the known ``electric-magnetic'' dualities.
We conjecture that this $SL(2,\bZ)$ transformation group, together 
with the standard
$SL(n,\bZ)$ rotations (which do not generally commute with it below 8
dimensions) and $\bZ_2$ parity transformations, 
generates all symmetries of M theory.

Our analysis is purely classical, assuming only that in the classical
low-energy limit M theory is correctly described by the supermembrane
action. The actual quantum theory may be a theory of strings for which
the membrane is the target space \km, a theory of strings for which
the fivebrane is the target space \dvv, or something completely
different that we have not yet been able to imagine. As shown in
\newhw, quantum corrections are definitely needed in order to
properly define M theory. At our present level of understanding, we
can only hope that in the quantum theory the membrane and wrapped
fivebrane will also be equivalent, and the duality we describe will
survive. The existence of a duality relating membranes and fivebranes
may suggest that they should be related already in the formulation of
the theory, as suggested in \dvv.
Quantum effects are believed to break the classical U duality
groups of supergravity to discrete subgroups \hulltowns, in a way
which has not yet been completely understood. We will not discuss this
issue here.

In section 2 we discuss the scalar-vector duality in the membrane of
M theory and its
relation to T duality in string theory.
In section 3 we analyze a particular
duality transformation in type II string theory compactified on a
torus, and see how it acts on the 11 dimensional fields.
In section 4 we show that this transformation exchanges the action of
the membrane of M theory (compactified on $\bT^3$) with the action of the
fivebrane wrapped around $\bT^3$.
In section 5 we show that from this 8
dimensional duality we can derive 
all of the known string dualities which
have a straightforward M theory interpretation. 

\newsec{T duality in M theory}

The simplest ``derivation'' of T duality in string theory comes from
looking at it as a scalar-scalar duality in the 1+1 dimensional
worldsheet (string theory T duality is described in \tdual\ 
and references within). 
If none of the background fields depend on one of the
spacetime dimensions, say $X_9$, then it enters into the worldsheet
lagrangian only through its derivative $\del_\alpha X_9$. We can replace
this derivative by a worldsheet gauge field $V_\alpha$, if we also add
another term to the string action ensuring that the associated gauge
field strength is zero. This term is just $\Lambda 
\epsilon^{\alpha \beta} \del_\alpha V_\beta$, where $\Lambda$ is a
Lagrange multiplier. By integrating out the Lagrange multiplier
$\Lambda$ we find that $V_\alpha$ is a total derivative, and we
return to the original action. If, however, we integrate over
$V_\alpha$ instead, we find a dual formulation of the action, in which
$\Lambda$ becomes a dynamical scalar field. We can only
perform this integration simply
if the derivatives appear in the action quadratically, so that
we should use the Polyakov form of the action and not the Nambu-Goto
form. By performing this sort of
1+1 dimensional scalar-scalar dualities we can derive any T duality
transformation. 

In M theory, strings generally arise by wrapping a membrane
around a compact dimension \memred. 
This suggests that T duality in M theory
should be a duality transformation in the worldvolume of the
membrane. Particularly, in 2+1 dimensions there is a duality
transformation transforming a scalar field (on which the action
depends only through its derivative) into a vector field, and this
goes over to the scalar-scalar duality described above when we
dimensionally reduce the membrane to the string. For instance, let us
describe the T duality of type IIA theory on a circle in M
theory. We begin with M theory on a torus, and perform a duality
transformation on the circle of the type IIA theory, exchanging
$\del_{\alpha} X_{9}$ by $V_{\alpha}$, adding a Lagrange multiplier
term $\epsilon^{\alpha \beta \gamma} \Lambda_{\alpha} \del_{\beta}
V_{\gamma}$, and integrating out $V_{\alpha}$. Again, we can do this
in a simple way only in a
formulation of the supermembrane action in which the derivatives
enter quadratically, and luckily such a formulation indeed exists \bst.
This transformation is, in fact, 
known to give the action of the D-2-brane \refs{\dfromm,\schmid}. 
Next, to go over to the
string theory, we dimensionally reduce this theory 
along the eleventh dimension $X_{10}$,
by setting one of the membrane coordinates $\xi_2$ to be exactly
proportional to $X_{10}$. The membrane gauge field $\Lambda_{\alpha}$
now becomes a gauge field on the string worldsheet and a scalar
$\Lambda_2$. The gauge field has no dynamical degrees of freedom (as
in the D-string), and the scalar becomes an additional, tenth,
dimension. It is easy to check that the metric of the ten new
scalars is just the T-dual of the original metric. 
Starting with the supermembrane action in 11 dimensions (whose
dimensional reduction gives the
type IIA string action \memred), we end up after the duality
with a type IIB string action \tdualII. Thus, T duality in
M theory is simply scalar-vector duality in the membrane worldvolume.

The procedure described above can easily be performed once, but
becomes more complicated when we try to perform it for more scalar
fields, which have couplings between them (through the metric or
3-form fields). Unlike the string action, the membrane action is
generally not quadratic in the fields $\del_{\alpha} X^{\mu}$.
Therefore, for general background fields (which we
always assume not to depend on the compact coordinates we want
to dualize), we can only perform explicitly 
two duality transformations. After
that, if the 3-form field corresponding to the 3 directions we want to
dualize does not vanish,
the action for the fields we want to integrate out includes higher
than quadratic terms, and we do not know how to integrate them
out in a simple way. Of course,
we can always leave the auxiliary fields in and be left with a 
more complicated description of the dual theory. Upon dimensional
reduction to string theory these problems disappear, since the string
action is always quadratic in the fields $\del_{\alpha} X^{\mu}$.

\newsec{T duality in 8 dimensions}

The ``single'' T duality transformation described above
is not really a symmetry of M
theory, since it exchanges one type of membrane theory (which has 
only
scalar fields) with a different type of theory (which has also a
vector field). This is not surprising, since this is not really a
symmetry in string theory as well, where it exchanges different types
of string theories. However, once we compactify two dimensions in
string theory, we have duality transformations which leave us in the
same string theory, and these should be genuine symmetries of M
theory (they should certainly be symmetries at least of eleven
dimensional supergravity). 
In this section we will analyze the simplest transformation of
this type, which is the
duality inverting the area of a torus in type IIA string
theory. Since we know how this duality acts on the fields of the
low-energy type IIA supergravity, we can find how it acts on the
fields of the 11 dimensional supergravity, because we know how
the two are related. As we will show,
this duality is actually an 8 dimensional electric-magnetic
duality in M theory, exchanging the 3-form field with its dual.
Then, in the next section,
we will show that in M theory the duality exchanges the membrane
action expressed in terms of the
original background fields, with the action of a fivebrane wrapped
around the 3-torus in the dual background fields.

We begin by finding the transformation of this duality on the fields
of the low-energy 11 dimensional supergravity.
Since we know how T duality acts on the low-energy fields and on
the D-branes in string theory, all we need is to translate the
transformation of these fields and $p$-branes to M theory. In fact,
the exact action of T duality on all RR fields has not been computed
as far as we know (for 9 dimensional T duality it is given in \sugra),
but we will use approximations in which the
transformations of these fields are simple.
For simplicity, we will begin by taking the type IIA string theory to
be compactified on a torus with a diagonal metric, with radii $r_8$
and $r_9$ in the string metric (we will denote string theory fields
and radii with small letters, and M theory fields and radii with
capital letters). We will also work only to leading order in the
off diagonal fields $G_{\mu i}$, $C_{\mu i j}$ and $C_{\mu \nu i}$,
where $\mu,\nu=0,\cdots,7$ and $i,j=8,9,10$ (this is the notation we
will generally use in this paper).
The exact expressions are known, at least for the NS-NS fields,
but they are much more cumbersome and do not seem to involve any new
issues. We will discuss here only the transformations of the
bosonic fields. The transformations of the fermionic fields are
related to these by supersymmetry.

T duality transformations on a torus act naturally on the K\"ahler
structure parameter $\tau = b_{89} + i r_8 r_9$ (which is often
denoted by $\rho$). The T duality group includes
$SL(2,\bZ)$ transformations of this parameter, and we will be
interested in the transformation taking $\tau \to -1/ \tau$, which
(for $b_{89}=0$) inverts both radii of the torus. In
a diagonal metric, this transformation takes $r_8$ to $r_8 / |\tau|$ and
$r_9$ to $r_9 / |\tau|$ (when $b_{89}=0$ we can regard this
transformation as a T duality on $r_8$, followed by a 
T duality on $r_9$, followed by a rotation
exchanging the two coordinates). The string coupling $\lambda$
transforms as $\lambda \to \lambda /
|\tau|$. Next, we should translate these results to M theory. The
string coupling is related to the radius $R_{10}$ of the eleventh
dimension by $\lambda = R_{10}^{3/2}$, while the relation between the
string theory and M theory metrics sets $r_8 = R_8 \sqrt{R_{10}}$ and
$r_9 = R_9 \sqrt{R_{10}}$. The tensor fields are related by $b_{\mu
\nu} = C_{\mu \nu (10)}$. Thus, in M theory $\tau$ is simply given by
$\tau = C_{89(10)} + i R_8 R_9 R_{10}$, 
with the same transformation law $\tau \to -1/\tau$ (as also noted by 
Sen \tsen), while the
11 dimensional radii transform as
\eqn\radiitrans{R_8 \to {R_8 \over |\tau|^{2/3}}  \qquad R_9 \to {R_9
\over |\tau|^{2/3}} \qquad R_{10} \to {R_{10} \over |\tau|^{2/3}}.}
The fact that these transformations are symmetrical in the three
compact dimensions suggests that this particular T duality
may indeed be given by a simple 
transformation in M theory. The other T duality
transformations are the shift in the $b_{89}=C_{89(10)}$ field, and the
$SL(2,\bZ)$ transformations of the complex structure, which are both
obviously expected to be symmetries in M theory as well.

Next, let us examine how the duality acts on the 8 dimensional
metric. The string metric $g_{\mu \nu}$
does not change under the duality (in the leading order approximation
we are working in), but the
radius of the eleventh dimension does change (by equation
\radiitrans). Therefore,
the 8 dimensional metric in 11 dimensional units changes by
$G_{\mu \nu} \to G_{\mu \nu} |\tau|^{2/3}$.

We will discuss the transformation of the off-diagonal metric elements
below, but first let us analyze the transformation of the 3-form
field, $C_{\mu \nu \lambda}$. In M theory this couples to the
membrane, so in the type IIA theory on a torus it couples to a 2-brane
which is not wrapped around any cycle of the torus. After the T duality,
this becomes a 4-brane which is wrapped around both cycles of the
torus, which in M theory is described by 
a fivebrane wrapped around $\bT^3$. In
11 dimensions the fivebrane couples to the dual $\tilde C$ of $C_{\mu \nu
\lambda}$. Thus, in 8 dimensions the fivebrane wrapped around $\bT^3$
couples to the 8 dimensional dual ${\tilde C}_{\mu \nu \lambda}$ of
$C_{\mu \nu \lambda}$ (we will discuss the exact definition of $\tilde
C$ in the next section). Therefore, the T duality transformation exchanges
the 8 dimensional 3-form field with its electric-magnetic dual, and in
this sense this duality is a membrane-membrane duality as originally
proposed by Townsend \memmem. By doing the transformation more
carefully we find that in fact it is given, to leading order, by
\eqn\ctrans{C_{\mu \nu \lambda} \to C_{89(10)} C_{\mu \nu \lambda} +
R_8 R_9 R_{10} {\tilde C}_{\mu \nu \lambda}.}
More general $SL(2,\bZ)$ transformations will mix all of the dyonic
membranes found in \dyonic.
As we will see below, other components of the 11 dimensional
3-form field do not transform into their duals, so that this
duality does not seem to be related (at least directly)
to an 11 dimensional
electric-magnetic (membrane-fivebrane) duality. In fact, we already
saw above that $C_{89(10)}$ has a simple transformation, which is not
related to electric-magnetic duality.

Let us now check the transformation of the off-diagonal components of
the 3-form field, starting with the fields $C_{\mu \nu i}$ (for
$i=8,9,10$). If $i=8$,
for instance, this field couples to a 2-brane wrapped around
$r_8$. When $b_{89}=0$, we can do a T duality transformation on $r_8$,
which turns this into an unwrapped 1-brane, then a T duality
transformation on $r_9$, which turns this into a 2-brane wrapped
around $r_9$, and finally a rotation exchanging the two circles of 
the torus, which returns this to a 2-brane wrapped around
$r_8$. Thus, $C_{\mu \nu 8}$ is actually invariant under the
duality. $C_{\mu \nu (10)}$ is just the 2-form field $b_{\mu \nu}$ of
the string theory, which is also invariant under the duality (to
leading order in the
off-diagonal background fields). Therefore, the
duality leaves the fields $C_{\mu \nu i}$ invariant (to leading order
in the off-diagonal fields).

The fields $C_{\mu i j}$, on the other hand, are not invariant. Let us
begin, for instance, with $C_{\mu 8 (10)}$. In the string theory this
is just $b_{\mu 8}$, which transforms under the T duality to
$(b_{89}b_{\mu 8}-g_{88}g_{\mu 9}) / |\tau|^2$.
The transformation of
$C_{\mu 9 (10)}$ is analogous. The other 8 dimensional vector field,
$C_{\mu 8 9}$, is part of the RR 3-form
in the type IIA string theory, which couples to a 2-brane wrapped
around both $r_8$ and $r_9$. Under T duality this becomes a 0-brane,
coupling to the RR gauge field $A_\mu$, which in 11 dimensional terms is
proportional to $G_{\mu (10)}$ (with the constant of proportionality
equal to $R_{10}^2$). In fact, $C_{\mu 8 9}$ mixes with $A_{\mu}
b_{89}$, and the actual transformation is slightly more
complicated.
Translating all this to 11 dimensions,
we find that the transformation
of these fields is given by
\eqn\transcmij{
C_{\mu i j} \to {1\over |\tau|^2} (-\epsilon^{ijk} 
G_{\mu k} R_i^2 R_j^2 - C_{89(10)} C_{\mu i j}).}
Analogously, one can compute the transformation of the fields $G_{\mu
i}$, and find that their transformation, in terms of 11 dimensional
fields, is given (to leading order in the off-diagonal fields) by
\eqn\transgmi{
G_{\mu i} \to {1\over |\tau|^{4/3}} (\half R_i^2
\epsilon^{ijk} C_{\mu jk} - C_{8 9 (10)} G_{\mu i}).}
Thus, the duality exchanges a membrane wrapped around two cycles of the
torus $\bT^3$ with a momentum mode around the third cycle, as in \tsen.
Note that by performing the T duality transformation twice, the 8
dimensional vector
fields change sign, so that we get (in 11 dimensional terms)
a parity transformation on the 3
compact dimensions, together with a change in the sign of the 3-form
field $C$. This can also be seen by the fact that $C_{\mu \nu
\lambda}$ transforms essentially
by electric-magnetic duality, which squares to $(-1)$ in 8 dimensions. 
All other fields are invariant under the double transformation.
This $\bZ_2$ is, of course, expected to be a symmetry of M
theory, like any change of sign in an odd number of dimensions and in
the $C$ field.

This completes the transformations of all fields in the eleven
dimensional low-energy supergravity theory, so let us summarize our
results. We have performed a T duality transformation in the type IIA
theory on a torus, involving the inversion of the K\"ahler
structure of the torus. Then, we translated this transformation to M
theory. Obviously this duality should be a symmetry of M theory as
well. Generalizing our previous results to a general metric on the
torus, we find that
the transformation of all bosonic fields of the 11 dimensional
supergravity is given, to
leading order in the off-diagonal fields, by :
\eqn\summary{\eqalign{
G_{ij} &\to {G_{ij} \over |\tau|^{4/3}} \cr
C_{89(10)} &\to -{C_{89(10)} \over |\tau|^2} \cr
C_{\mu \nu \lambda} &\to C_{89(10)} C_{\mu \nu \lambda} + \sqrt{\dg}
{\tilde C}_{\mu \nu \lambda} \cr
C_{\mu \nu i} &\to C_{\mu \nu i} \cr
C_{\mu i j} &\to {1\over |\tau|^2} (-\epsilon^{i'j'k'} 
G_{ii'} G_{jj'} G_{\mu k'} - C_{89(10)} C_{\mu i j}) \cr
G_{\mu \nu} &\to G_{\mu \nu} |\tau|^{2/3} \cr
G_{\mu i} &\to {1\over |\tau|^{4/3}} (\half
\epsilon^{i'j'k'} G_{ii'} C_{\mu j'k'} - C_{8 9 (10)} G_{\mu i}). \cr }}

\newsec{Membrane-membrane duality in 8 dimensions}

We would now like to identify the symmetry described in the previous
section in M theory. For a membrane wrapped around any cycle of the
torus, it is obviously just a T duality. This is clear for a membrane
wrapped around $x_{10}$ from our definition of the transformation, but
since our results are symmetric it is true for any wrapped
membrane. Thus, we should only identify how an unwrapped membrane
transforms. The transformation of the 3-form field suggests that it
should transform into a fivebrane wrapped around $\bT^3$.
In this section we will
examine how the transformations \summary\ indeed relate the action of
the membrane with the action of the completely wrapped fivebrane.
Fivebranes which are wrapped around less than three cycles of the
3-torus transform into themselves according to \summary, and we will
not discuss them further.

First, let us write down the action of the membrane.
We will use the Howe-Tucker form of the
supermembrane action \bst,
which has an auxiliary metric $\gamma_{\alpha \beta}$ on the
worldvolume of the membrane, and write only the purely bosonic terms
throughout this section (the addition of the fermionic terms is not
expected to change our results, since they should be determined by
supersymmetry). 
Separating the compact and non-compact directions, this action is
\eqn\maction{
\eqalign{S_M = \int d^3 \xi \{ -\half \sqrt{-\gamma}
& \left[ \gamma^{\alpha \beta} (\del_{\alpha} X^{\mu} \del_{\beta} X^{\nu}
G_{\mu \nu} + 2 \del_{\alpha} X^{\mu} \del_{\beta} X^i G_{\mu i} +
\del_{\alpha} X^i \del_{\beta} X^j G_{ij}) - 1 \right] \cr
-{1\over 6} \epsilon^{\alpha \beta \gamma} (&C_{\mu \nu \rho}
\del_{\alpha} X^{\mu} \del_{\beta} X^{\nu} \del_{\gamma} X^{\rho} +
3 C_{\mu \nu i} \del_{\alpha} X^{\mu} \del_{\beta} X^{\nu}
\del_{\gamma} X^i \cr &+ 3 C_{\mu i j} \del_{\alpha} X^{\mu}
\del_{\beta} X^i \del_{\gamma} X^j +
C_{ijk} \del_{\alpha} X^i \del_{\beta} X^j \del_{\gamma} X^k) \}. \cr}}
Note that throughout this paper we use conventions in which epsilon
symbols with upper indices are equal to $\pm 1$.

We would like to compare this result with the action for the M theory
fivebrane wrapped around the 3-torus. However,
we do not know how to write an action for this fivebrane, due to
the existence of a self-dual 2-form field $B_{ab}$
in its worldvolume\foot{We
thank P. K. Townsend for discussions on this issue.}.
Townsend has suggested \dfromm\ an action for the fivebrane 
(at least in the low-energy limit) which is gauge invariant, and
may describe correctly at least some properties of the fivebrane
if the self-duality condition is added by hand to the equations of motion. 
It is given, in 11 dimensions, by
\eqn\faction{\eqalign{S_F^{(0)}=-\half \int d^6 \xi \sqrt{-\gamma} [
\gamma^{ab} G_{MN} \del_a X^M & \del_b X^N - 4 \cr
+\half \gamma^{ad} \gamma^{be} \gamma^{cf} & (F_{abc} - C_{MNP}
\del_a X^M \del_b X^N \del_c X^P) \cr & (F_{def} - 
C_{M'N'P'} \del_d X^{M'} \del_e X^{N'} \del_f 
X^{P'})], \cr }}
where $F_{abc}$ is the field strength associated with the 2-form field
$B_{ab}$. This is gauge invariant if the gauge transformation 
$C_{MNP} \to C_{MNP}
+ \del_{[M} \Lambda_{NP]}$ is accompanied by a
shift 
in the 2-form field of the fivebrane,
$B_{ab} \to B_{ab} + \Lambda_{NP} \del_a X^N \del_b
X^P$. 
The self-duality condition must also be changed from stating that 
$F_{abc}$ is
self-dual to stating that $F_{abc}-{\hat C}_{abc}$ is self-dual
(where $\hat C$ is
the pullback of $C$ to the fivebrane worldvolume, ${\hat C}_{abc} =
C_{MNP} \del_a X^M \del_b X^N \del_c
X^P$), since only this combination is gauge invariant.
The equations of motion arising from this action 
are consistent with the self-duality condition after we add an
additional term to the action as described below.
Upon dimensionally
reducing three dimensions from the fivebrane, the self-duality conditions
may be trivially resolved (as described below). Thus, we can hope that
this action, together with the correction described below,
may indeed be
a correct description for the wrapped fivebrane, even though in 11
dimensions the
self-duality condition has to be added to it by hand.

We expect the fivebrane action to also include another term,
describing the coupling to the dual $\tilde C$ of the 3-form field. 
Before
discussing this we should define exactly what we mean by $\tilde
C$, since the equation of motion of the $C$ field in 11 dimensional
supergravity is $d(*dC)=dC \wedge dC$ (we ignore numerical constants
in this paragraph), and we cannot in general define
a field $\tilde C$ by $*dC = d{\tilde C}$. However, since
$d(*dC-C\wedge dC)=0$, we can define $d{\tilde C}=*dC - C \wedge dC$,
and this is the definition we will be using for $\tilde C$.
This definition means that $\tilde C$ is not invariant under gauge
transformations of $C$. If $C \to C + d\Lambda$, $\tilde C$
transforms by ${\tilde C} \to {\tilde C} + d\Lambda \wedge C$. Thus,
we cannot write a Lagrangian for the fivebrane with a term
$\int d^6 \xi {\hat{\tilde C}}$ (where $\hat{\tilde C}$ 
is the pullback of
the 6-form field $\tilde C$ to the worldvolume of the fivebrane), 
as we would like to, since this is not
gauge invariant. We cannot fix this by a term proportional to
${\hat C} \wedge {\hat C}$, since this vanishes. Instead,
the only gauge invariant lagrangian which can write seems to be
$S_F^{(1)} = \int d^6 \xi ({\hat{\tilde C}} - F \wedge {\hat C})$, and,
therefore, the complete lagrangian we will use for the fivebrane is
$S_F = S_F^{(0)} - S_F^{(1)}$. Fortunately, 
the equations of motion derived from this action are consistent
with the self-duality condition we described above, and $S_F$ seems
to be a consistent action. The 11 dimensional
supergravity theory also seems to be
consistent with this form of the fivebrane action, as discussed in
\witorbs.

Let us now discuss the field content we obtain 
when reducing this action to the action of a membrane in
8 dimensions. To perform the dimensional reduction we choose three 
of the
fivebrane coordinates to equal the coordinates of the torus, $\xi_3 =
X_8, \xi_4 = X_9, \xi_5 = X_{10}$, and then we can perform the
integration over these coordinates (since the background fields do not
depend on them). We can do this in a simple way only if $G_{\mu i}=0$,
and we will assume this from here on in the fivebrane theory
(according to \summary, when $C_{89(10)}=0$
this is dual to assuming that in the membrane action $C_{\mu
i j} = 0$).
We will be left with a membrane action with 8 scalar
fields $X_{\mu}$ ($\mu=0,\cdots,7$), and with what remains of the
self-dual 2-form. As discussed above,
the self-duality condition on the 3-form field strength is
\eqn\selfdual{F_{abc}-{\hat C}_{abc}={1\over 6\sqrt{-\gamma}} \gamma_{aa'}
\gamma_{bb'} \gamma_{cc'} \epsilon^{a'b'c'def} (F_{def}-{\hat
C}_{def}).} 
When none of the fields
depend on the last 3 coordinates of the fivebrane, this
expression simplifies considerably. For $a,b,c=0,1,2$ we find
that $(F_{012} - {\hat C}_{012}) \propto (F_{345} - {\hat C}_{345}) =
-{\hat C}_{345}$, since $F_{345}=0$. Thus,
$F_{012}$ is no longer an independent dynamical field. For
$a,b,c=0,1,3$ (for instance), we find that
$(\del_0 B_{13} - \del_1 B_{03} - {\hat C}_{013}) \propto (\del_2
B_{54} - {\hat C}_{254})$
(assuming for the moment that the metric $\gamma$ is diagonal). Hence,
the scalar field $B_{45}$ is determined in terms of the vector
field $B_{\alpha 3}$ (and vice versa). 
The analysis of the other components of the
self-duality equations is analogous, and we find that we can remain
either with
three independent vector fields $B_{\alpha a}$
(where $\alpha=0,1,2$ is a membrane worldvolume index and $a=3,4,5$) 
in the membrane worldvolume, or with three independent scalar fields
$B_{ab}$ ($a,b=3,4,5$). We will choose to remain with the scalar
fields, since we will show that these may be identified in a simple
way with the scalar fields in the membrane theory. Leaving the vector
fields would lead to an action which is related to the scalar action
by a triple scalar-vector duality of the type described in section 2.

After the reduction, we can just throw away the terms
in the action involving $F_{\alpha \beta a}$ (from here on $\alpha,\beta,
\gamma=0,1,2$ and $a,b,c=3,4,5$), 
since the equations of motion of these vector fields just give, when
using the self-duality equation, the Bianchi identity for the scalar
fields (which is $\epsilon^{\alpha \beta \gamma} \epsilon^{abc}
\del_{\alpha} \del_{\beta} B_{ab} = 0$). Thus, retaining these terms
does not add any new information.
In the same way we can throw away the terms
involving $F_{012}$, since they also give trivial equations of motion
(when using the self-duality condition). Hence, the only terms
involving $F$ which remain in the action are those involving
$F_{\alpha a b}$.
The field content we find is, therefore, the same as the field content
in the membrane action \maction \memmem. 

We would now like to identify the terms in the action \maction\ with
the terms in the reduction of \faction, according to the
transformation \summary.
Let us begin
with the metric field $G_{\mu \nu}$. In both \maction\ and \faction\
this field appears canonically, but we should recall that we have
(implicitly) set the tensions of the membrane and the fivebrane to one
in the above formulas, and these determine the length scale by which
the metric is measured. In 11 dimensional units, the tensions are
related by \schwarz\ 
$T_5 = (T_2)^2$ (up to numerical factors which we suppress). When
$C_{89(10)}=0$,
the tension of the wrapped fivebrane is $T_5 \sqrt{\dg}$, and we
see that in units in which the membrane tension is one, it is just
the volume of the 3-torus,
$\sqrt{\dg}$. Thus, the scales of the two theories are related
by a factor of $(\dg)^{1/6}$. Since the metric has
dimensions of length squared, this gives exactly the
relation in \summary\ between the 8 dimensional metrics, as expected
(recall that generally $\tau = C_{89(10)} + i \sqrt{\dg}$). For a
non-vanishing $C_{89(10)}$ field, the relation is more complicated due
to the presence of a $C_{89(10)}^2$ term in the wrapped fivebrane
action. To simplify the equations we will assume from here on that
$C_{89(10)}=0$. 

Next,
we can easily compare the terms which are linear in
the 8 dimensional 3-form fields. In
the wrapped fivebrane action the $\tilde C$ term in the action becomes
$-{1\over 6}\epsilon^{\alpha \beta \gamma} {\tilde C}_{\mu \nu \rho}
\del_{\alpha} X^{\mu} \del_{\beta} X^{\nu} \del_{\gamma} X^{\rho}$. 
Identifying this with the 3-form term in \faction, using the
relation described above between the membrane and fivebrane metrics, we
find exactly the relation \summary\ between the background fields in
the 2 actions, as desired.

The relation between the other terms in the two actions is
slightly more complicated. We claim that the actions are related by
the transformation \summary\ if we identify the scalars
$\half \epsilon^{abc} B_{ab}$
($a,b,c=3,4,5$) in the
fivebrane action with 
$X^{c+5}$ in the membrane action \maction. Note that this
identification exchanges the gauge transformation of the 3-form which
shifts $B_{ab}$ with
the isometry corresponding to a shift in $X^i$, as is also evident
from \summary.
In order to perform
the comparison we should replace the metric $\gamma^{ab}$ in the
internal directions by its classical value, according to the equations
of motion. In the absence of the second term in the action \faction,
this is just $\gamma_{ab} = G_{(a+5)(b+5)}$, but generally there are
corrections to this, arising from the $(F-{\hat C})^2$ term. 
In the leading
order approximation in which we analyzed the transformation in section
3, it is justified to ignore these corrections, since they are of the
same order as the terms which we ignored. Plugging in this solution,
and the relation between $B_{ab}$ and $X^i$,
we find that the quadratic term in the $B_{ab}$ fields becomes exactly
the term $-\half \sqrt{-\gamma} \gamma^{\alpha \beta} \del_{\alpha}
X^i \del_{\beta} X^j G_{ij}$ in the membrane action. The term in
\faction\ linear in $B_{ab}$ becomes, using \summary, 
the term $-\sqrt{-\gamma}
\del_{\alpha} X^{\mu} \del_{\beta} X^i G_{\mu i}$ in the membrane 
action, while the term
linear in $B_{ab}$ from the $F\wedge {\hat C}$ term becomes just
$-{1\over 2}\epsilon^{\alpha \beta \gamma} C_{\mu \nu i} \del_{\alpha}
X^{\mu} \del_{\beta} X^{\nu} \del_{\gamma} X^i$, which we equate with
the same term in \maction\ (since $C_{\mu \nu i}$ is invariant under
the duality).
Since we have taken $G_{\mu i}=0$ in the fivebrane action and ignored
all higher order terms, these are the only terms we can compare.
Presumably, when doing
the exact transformation instead of \summary\ and plugging in the
exact solution for $\gamma_{ab}$, the other terms will match as well. 

We
conclude that, to the extent that we have checked it, the membrane action
in the original background seems to be the same as the
wrapped fivebrane action in the dual background given by \summary.
At least in the approximation we used,
a simple form of
the fivebrane action, supplemented by the self-duality condition,
seems to describe the fivebrane worldvolume
theory in a way consistent with the duality.

Let us end this section with a comment about the relevance of this
duality to the issue of the length scales in M theory.
When $C_{89(10)}=0$, the duality transformation inverts the 3-volume
of the torus $V=\sqrt{\dg}$. In M theory there does not seem to
exist a minimal length scale, at least classically.
Taking M theory on a radius much
smaller than the 11 dimensional Planck scale just leads to a weakly
coupled string theory. However, the existence of the duality
transformation we described suggests that the 11 dimensional
Planck scale may serve as a minimal
3-volume scale in M theory, since M theory on a 3-torus of volume $V$ is
equivalent to M theory on a 3-torus of volume $1/V$. This is perhaps
natural in some sense if M theory is indeed a theory of membranes.
Unfortunately,
this interpretation is not clear-cut, since the duality also changes the
metric in the remaining 8 directions, unlike T duality in string theory.
In any case, we hope that understanding this duality symmetry
may shed some light on the problem of understanding the length scales
in string theory and in M theory \shenker.

\newsec{String dualities from membrane-membrane duality}

In this section we will derive many of the known string dualities from
the M theory duality described in the previous two sections. Our general
strategy will be to take the original M theory, perhaps compactified
along more directions, and orbifold it by some discrete
symmetries. The relation \summary\ between the variables of the
original and of the dual theories will allow us to identify these
discrete symmetries in the dual theory, so that we will know how to
perform the orbifold also in the dual theory. In this way we will
obtain a duality transformation relating
two, generally different, orbifolds of M theory.

First, we can easily get the T duality of the heterotic string on a
torus from the duality transformation of the previous section, by
simply orbifolding by $x_{10} \leftrightarrow -x_{10}$, together with
$C \leftrightarrow -C$ (in the membrane worldvolume this involves a
parity transformation as well). This transformation is invariant under the
duality. Thus, on both sides we get the heterotic $E_8 \times E_8$
string on $\bT^2$, and the relation between them is just the usual T
duality of this theory on the torus.

Less trivial dualities arise if we add more compact dimensions and
orbifold by different symmetries. Without orbifolding, or by performing
only the orbifold of the previous paragraph, we can of course get more
complicated T duality transformations for the type II and heterotic
strings. The orbifolds we discuss below are all of type 2(b)
in the classification proposed by Sen \newsen\ of orbifolds and
dualities. Therefore, we expect the duality to commute with the
orbifolding in these cases, and indeed we will always find a pair of
dual theories.
Let us begin by adding another compact dimension $x_7$, and
orbifolding by the symmetry which changes the sign of all four compact
dimensions. In the original theory we would thus get M theory on
$\bT^4/\bZ_2$, which is an orbifold limit of K3. Using \summary\ (or
working directly in the worldvolume of the membrane), we
find that in the dual theory this symmetry changes the sign of $x_7$
and of the 3-form field $C$, and we get the heterotic string
theory on $\bT^3$. Thus, the 8 dimensional membrane-membrane duality
naturally leads to a duality between M theory on K3 and the heterotic
string theory on $\bT^3$ \witten. 

By adding another compact dimension we can easily
derive from this the duality between type IIA theory on K3 and
the heterotic string theory on $\bT^4$. If we then perform another
orbifold, by the symmetry which changes the sign of $x_6$ and $C$ in
the original theory, and changes the sign of $x_6,x_8,x_9,x_{10}$ in
the dual theory, this leads to the heterotic-heterotic duality of the
heterotic string theory on K3 \hethet. 
One can easily check that the transformation
rules \summary\ indeed exchange the fivebrane wrapped around K3 in the
original theory with the membrane wrapped around
$x_7$ in the dual theory, as expected.
Thus, these string-string dualities may also be
derived directly from the membrane-membrane duality of M theory
(at least in the orbifold limit of K3).
We do not yet know how to
define (from first principles) the twisted sectors in orbifolds 
of M theory. However,
once we know how to do this, the description given above of the duality
transformation may allow us to find the exact relation between the
twisted fields before and after the duality transformation. We do not
know how to generalize these dualities to generic K3 manifolds in our
framework, but presumably this should also be possible.

Other dualities in string theory involve the type IIB string theory on
various backgrounds (including ``F theory'' backgrounds \ftheory) 
which do not
include a circle. The description of these in M theory seems to 
be singular,
because we must take the limit in which the area of a torus goes 
to zero in order
to get the type IIB theory in ten dimensions \schwarz. 
Since there is no clear evidence for
a minimal length scale in M theory, it is not clear that this limit
is indeed singular. In any case, we can try
to derive these dualities upon compactifying on an additional $\bS^1$,
when the type IIB theory is identical to the type IIA theory. For
instance, let us derive the duality between M theory on
$(\bS^1)^5/\bZ_2$ and type IIB theory on K3 \refs{\dmorbs,\witorbs} 
when compactified on an
additional circle $\bS^1$. We add three more compact dimensions, and
orbifold the original theory by the symmetry which changes the sign of
$x_5,x_6,x_7,x_8,x_9$ and $C$, leading to M theory on $(\bS^1)^5/\bZ_2
\times \bS^1$. In the dual theory, we find that this symmetry
corresponds to changing the sign of $x_5,x_6,x_7$ and $x_{10}$,
leading to the M theory on $\bT^4/\bZ_2 \times \bT^2$, which is an
orbifold limit of the type II theory on $K3 \times \bS^1$. Thus, this
duality also arises from the membrane-membrane duality in 8
dimensions, but in order to get it without the additional $\bS^1$ we
must take the limit $R_{10} \to \infty$ (in the original theory). 
Note that if we do not add an
additional $\bS^1$, and divide the original theory by the symmetry
changing the sign of $x_6,x_7,x_8,x_9,x_{10}$ and $C$, we find the
same symmetry in the dual theory, so this is just a T
duality of M theory on $(\bS^1)^5 / \bZ_2$. Translating this T
duality to the type IIB theory on $\bT^4/\bZ_2$, 
we find that it is just the
exchange of two of the circles in $\bT^4/\bZ_2$. 

\centerline{ }
\centerline{\bf Acknowledgments}

I would like to thank everyone at the New High Energy Theory Center
at Rutgers University for their hospitality and for creating a
stimulating environment during the course of this work. I would
also like to thank T. Banks, M. Berkooz, O. Ganor, N. Seiberg, 
S. H. Shenker and especially J. Sonnenschein, P. K. Townsend 
and S. Yankielowicz for useful discussions.

\listrefs

\end